# A Singularity in the First-order PY Equation for a Square Well Fluid

**Sheldon Shanack**
Occidental College Physics Dept.
1600 Campus Road, Box M-60
Los Angeles, CA 90041
E-mail: shanack@oxy.edu

Rodney Varley
Hunter College Dept. of Physics and Astronomy
695 Park Avenue, 1225 Hunter North
New York, NY 10065
E-mail: Rodney.Varley@hunter.cuny.edu

**Abstract.** It is shown that a nearest neighbor Square Well (SW) potential leads to singular behavior in first order. A solution of the first-order perturbative *PY* equation for an attractive nearest neighbor square well of width less than the core diameter reveals singular behavior. Neither the width of the well nor the depth of the well affects the location of the singularity in the inverse isothermal compressibility. The analysis is carried out in one dimension.

## 1. Introduction

One of the more reliable approximate theories of pair interacting bulk fluids in thermal equilibrium is given by the *PY* equation [1]. This has been solved in closed analytic form for only a few special models—hard cores [2]—also with "glue" [3]—which nonetheless have served as starting points for a number of modern treatments of more realistic fluids. Any extra-core potential is in some sense to be regarded as a perturbation, which may be of the "solved" reference fluid or of the *PY* equation itself. The latter tactic may seem to have the quality of self-consistency. However, it cannot be pushed too far, and in this paper we will show by application to very much of a toy model that it can generate singular behavior even at the level of first order perturbation, suitably defined.

The toy model we have in mind is that of 1-dimensional hard rods with a square well attraction appended:

$$V(x) = \begin{cases} \infty & a > x > 0 \\ -V_1 & a + \Delta a > x > a\ , \qquad \Delta a < a \\ 0 & x > a + \Delta a \end{cases} \tag{1}$$

where $a$ is the diameter of the core, $\Delta a$ is the well width, and $V_1$ is the well depth. The energy of the attractive interaction is $-V_1$.

As a one-dimensional system with finite range forces, the thermodynamics, if exact, will of course be well-behaved. In fact, standard solution in an isobaric ensemble [4] gives the explicit result ($\beta$ is reciprocal temperature, $\mu$ chemical potential, and $\ell$ specific volume):

$$\begin{aligned} \beta\mu &= -\ell n \int_0^\infty e^{-\beta P x}\, e^{-\beta V(x)}\, dx \\ &= \ell n\, \beta P + \beta P a - \beta V_1 - \ell n \left(1 + \left(1 - e^{-\beta V_1}\right) e^{-\beta P \Delta a}\right) \\ \ell\ &= \frac{\partial \beta\mu}{\partial \beta P} = \frac{1}{\beta P} + a + \frac{\Delta a \left(1 - e^{-\beta V_1}\right) e^{-\beta P \Delta a}}{1 + \left(1 - e^{-\beta V_1}\right) e^{-\beta P \Delta a}} \end{aligned}$$

A convenient quantity to focus on is the inverse compressibility

$$\beta K_T^{-1} = -\ell / \left(\partial \ell / \partial \beta P\right).$$

Of course, as the attractive Boltzmann factor $e^{\beta V_1}$ increases, the system will try to collapse, but cannot at fixed uniform density. What we will show is that, perhaps unexpectedly, there results a singularity in $\beta K_T^{-1}$ at a characteristic density, sensibly independent of the values of $V_1$, $\Delta a$. This suggests that there is a more complex model hovering in the background for which the singular behavior is legitimate.

An approximate solution for a one-dimensional *SW* fluid can be obtained by solving the first-order *PY* perturbative equation, obtained as follows:

First, write $V(x)$ as a sum of two parts, the repulsive hard core $V_0(x)$ plus the attractive potential $V_1(x)$,

$$V(x) = V_0(x) + V_1(x)$$

where

$$\begin{aligned} V_0(x) &= \infty, \qquad 0 < |x| < a \\ &= 0, \qquad |x| > a \end{aligned} \tag{2a}$$

also

$$\begin{aligned} V_1(x) &= 0, \qquad 0 < |x| < a \\ &\neq 0, \qquad a < |x| < a + \Delta a \end{aligned} \tag{2b}$$

The quantity $e_1(x)$ can be regarded as a perturbation on $e_0(x) = \exp(-\beta V_0(x))$, where $e_1(x) = e(x) - e_0(x)$, or

$$\begin{aligned} &e_1(x) = e_0(x) \left[\exp(-\beta V_1(x)) - 1\right] \\ &\text{with } \beta = (kT)^{-1}. \end{aligned} \tag{3}$$

The pair correlation function, $g(x)$, and the direct correlation function, $c(x)$, can then be expressed to first order in this perturbation scheme as

$$\begin{aligned} g(x) &= g_0(x) + g_1(x) \\ c(x) &= c_0(x) + c_1(x) \end{aligned} \tag{4}$$

The perturbative PY equation can be written in terms of $c_1(x)$ and $g_1(x)$. It is advantageous, however, to write the perturbative PY in terms of the direct correlation function $b_1(x)$ only, where

$$\begin{aligned} b_1(x) &= \hat{c}_1(x)\ , \qquad (|x| < a) \\ b_1(x) &= 0\ , \qquad (|x| > a) \end{aligned} \tag{5}$$

clearly,

$$\begin{aligned} c_1(x) &= \hat{c}_1(x) + \check{c}_1(x) \\ &= b_1(x) + g_0\, e_1(x) \end{aligned} \tag{6}$$

where $(\wedge)$ and $(\vee)$ denote inside and outside the core. The PY equation, perturbed to first-order, can be expressed in terms of $c_1(x)$ and $g_1(x)$ as

$$g_1(x) - c_1(x) = nc_0 * g_1 + n\, c_1 * (g_0 - 1)$$

$$\text{and} \qquad c_1(x) = g_0(x)\, e_1(x)\ , \quad |x| > a \tag{7}$$

where $n$ is the number density.

A detailed solution of equation (7) may be found in
`http://www.nyu.edu/pages/cimslibrary/, New York University Technical Report NYO-1480-151, 1970 .` [5]

The first-order direct correlation function inside the core, $b_1(x)$, is found to be [5]

$$\begin{aligned} b_1(x) &= 2n^2 Q^2 \int_0^x (x-\zeta)\, \Delta(\zeta)\, d\zeta - 2n^2 Q^2 \int_a^{a-x} (a - x - \zeta)\, \Delta(\zeta)\, d\zeta \\ &\quad -2n\, Q \int_a^{a-x} \Delta(\zeta)\, d\zeta - \Delta(x) + \\ &\quad + B_0 + B_1\, x + B_2\, x^2 + B_3\, x^3, \qquad (0 < x < a). \end{aligned} \tag{8}$$

where [5]:

$$\Delta(x) = \int_{-\infty}^{\infty} \Gamma\,(x - x')\, g_0(x')\, e_1(x')\, dx'. \tag{9}$$

## 2. Solution of Perturbative PY Equation for a Square-Well Fluid

We now calculate $b_1(x)$ from equations (8) and (9) for a SW potential. $\Delta(x)$ will be a continuous function of $x$ for $0 < x < a$, but will have different functional forms for each of the three intervals $0 < x < \Delta a$, $\Delta a < x < a - \Delta a$, and $a - \Delta a < x < a$, such that $0 < \Delta a < \frac{a}{2}$. (Similar considerations obtain for $\Delta a > a/2$). Thus, there will be different forms for $b_1(x)$ for each of the three intervals I, II, and III. From equation (8), $b_1(x)$ may be found for $(0 < x < a)$. In Region I $(0 < x < \Delta a)$, the first-order direct correlation function, denoted as $b_1^I(x)$, may be shown to be [5]

$$b_1^I(x) = \sum_{j=0}^{3} (B_j + \tilde{\nu}_j)\, X^j - Qe^{-nQX} - Qe^{-2nQ\,\Delta a} \cdot e^{nQX}. \tag{10}$$

In Region II $(\Delta a < x < a - \Delta a)$, it may be shown that

$$b_1^{II}(x) = \sum_{j=0}^{3} (B_j + \nu_j)\, X^j \tag{11}$$

and in Region III $(a - \Delta a < x < a)$

$$b_1^{III}(x) = \sum_{j=0}^{3} \left(B_j + \tilde{\tilde{\nu}}_j\right) X^j + A_3\, e^{nQ(x-a)} + A_4\, e^{-nQ(x-a)} + A_5 \left(x + \frac{4}{nQ}\right) e^{-nQ(x-a)} \tag{12}$$

In terms of the direct correlation function, [6]

$$\frac{\partial}{\partial n}\,\beta P = \frac{\partial}{\partial n}\,\beta\, P_0 - 2n \int_0^\infty \hat{c}_1\, dx - 2n \int_0^\infty \check{c}_1\, dx$$
$$\text{where } \frac{\partial}{\partial n}\,\beta\, P_0 = -c_0(0). \tag{13}$$

For a SW potential,

$$\frac{\partial}{\partial n}\,\beta P = Q^2 - 2e_1 \left(1 - e^{-nQ\,\Delta a}\right) +$$
$$+ 2ne_1 \left[\int_0^{\Delta a} b_1^I(x)\, dx + \int_{\Delta a}^{a-\Delta a} b_1^{II}(x)\, dx + \int_{a-\Delta a}^{a} b_1^{III}(x)\, dx\right]$$
$$\text{where } Q = (1 - na)^{-1}, \qquad \text{and} \qquad e_1 = \left(e^{-\beta V_1} - 1\right). \tag{14}$$

The SW perturbative solution may be conveniently written in terms of the isothermal compressibility, $K_T$, as (with $\ell = 1/n$):

$$\beta K_T^{-1} = \frac{Q^2}{\ell} - \frac{2e_1}{\ell} \left(1 - e^{-\frac{Q}{\ell}\Delta a}\right) +$$
$$+\frac{2}{\ell^2}\, e_1 \left[\int_0^{\Delta a} b_1^I(x)\, dx + \int_{\Delta a}^{a-\Delta a} b_1^{II}(x)\, dx + \int_{a-\Delta a}^{a} b_1^{III}(x)\, dx\right]. \tag{15}$$

From equations (10)–(12),

$$\int_0^{\Delta a} b_1^I(x)\, dx = \sum_{j=0}^{3} (B_j + \tilde{\nu}_j)\, \frac{(\Delta a)^{j+1}}{(j+1)} +$$
$$+\frac{1}{n} \left(e^{-nQ\,\Delta a} - 1\right) - \frac{1}{n}\, e^{-2nQ\Delta a} \left(e^{nQ\Delta a} - 1\right) \tag{16a}$$

$$\int_{\Delta a}^{a-\Delta a} b_1^{II}(x)\,dx = \sum_{j=0}^{3} \frac{(B_j+\nu_j)}{(j+1)} \left[(a-\Delta a)^{j+1} - (\Delta a)^{j+1}\right] \tag{16b}$$

$$\begin{aligned}\int_{a-\Delta a}^{a} b_1^{III}(x)\,dx = &\sum_{j=0}^{3} \frac{\left(B_j+\tilde{\tilde{\nu}}_j\right)}{(j+1)} \left[a^{j+1} - (a-\Delta a)^{j+1}\right] + \\ &+ \frac{A_3}{nQ}\left(1-e^{-nQ\Delta a}\right) - \frac{A_4}{nQ}\left(1-e^{nQ\Delta a}\right) \\ &- \frac{A_5}{(nQ)^2}\left[(Q+4) - e^{nQ\Delta a}\left(5+nQ\left(a-\Delta a\right)\right)\right].\end{aligned} \tag{16c}$$

The symbols which appear are defined in equation (17)

$$\begin{aligned}
K &= an^2Q^2\left(6-6an+\frac{10}{3}\,a^2 n^2 - a^3 n^3\right)\\
C_1 &= \left\{\left(6an\,Q^2+6Q-\frac{2K}{n}\right)+\left[-6n\left(a+\Delta a\right)Q^2-6Q+\frac{2K}{n}\right]e^{-nQ\,\Delta a}\right\}\\
C_2 &= 2n\,Q^2\\
\overline{B}_1 &= \left\{2\left[(1+2an)\,Q^2+2Q-\frac{K}{n}\right]-\right.\\
&\qquad \left.-2\left[(1+2n\left(a+\Delta a\right))\,Q^2+2Q-\frac{K}{n}\right]e^{-nQ\,\Delta a}\right\}\\
\overline{B}_2 &= 2n\,Q^2\left(1-e^{-nQ\,\Delta a}\right)\\
A_1 &= \left\{\left[2\left(1+2an\right)Q^2+4Q-\frac{2K}{n}\right]+\right.\\
&\qquad \left.+\left[-\left(2+\frac{11}{2}\,an+\frac{5}{2}\,n\,\Delta a\right)Q^2-\frac{5}{2}\,Q+\frac{2K}{n}\right]e^{-nQ\,\Delta a}\right\}\\
A_2 &= \frac{1}{2}\,n\,Q^2\left(4-e^{-n\,Q\,\Delta a}\right)\\
A_3 &= Q/4\\
A_4 &= \left\{\left[\frac{1}{2}\,n\left(\Delta a-a\right)Q^2-\frac{7}{4}\,Q\right]e^{-2n\,Q\,\Delta a}\right\}\\
A_5 &= \frac{1}{2}\,n\,Q^2\,e^{-2n\,Q\,\Delta a}\ .
\end{aligned} \tag{17}$$

Thus,

$$\begin{aligned}
&\left[\int_0^{\Delta a} b_1^{I}(x)\,dx + \int_{\Delta a}^{a-\Delta a} b_1^{II}(x)\,dx + \int_{a-\Delta a}^{a} b_1^{III}(x)\,dx\right] = \\
&\sum_{j=0}^{3}\frac{B_j}{(j+1)}\,a^{j+1} + \sum_{j=0}^{3}\frac{\tilde{\nu}_j}{(j+1)}\,(\Delta a)^{j+1} + \\
&+\sum_{j=0}^{3}\frac{\nu_j}{(j+1)}\left[(a-\Delta a)^{j+1}-(\Delta a)^{j+1}\right]+\sum_{j=0}^{3}\frac{\tilde{\nu}_j}{j+1}\left[a^{j+1}-(a-\Delta a)^{j+1}\right]\\
&+\frac{1}{2n}\,e^{-nQ\Delta a}+\left[\frac{1}{4n}-\frac{1}{2}\,(\Delta a)\,Q\right]e^{-2nQ\Delta a}-\frac{3}{4n}\,.
\end{aligned} \tag{18}$$

where

$$\nu_3 = \frac{4}{3}\,n^3\,Q^4\left(1-e^{-nQ\Delta a}\right)$$

$$\nu_2 = -2n^2 Q^4 \left(1 - e^{-nQ\Delta a}\right)$$
$$\nu_1 = Q\left[\frac{11}{2}\, nQ + 4n\,(1+3an)\,Q^2 - 4K\right]$$
$$+ e^{-nQ\Delta a}\left\{(\Delta a)^2\left(-\frac{7}{2}\, n^3 Q^4\right) + (\Delta a)\left(-11n^2\, Q^3\right) + \right.$$
$$\left. + Q\left[-5n\, Q^2 - 11an^2 Q^2 - 2nQ + 4K\right]\right\}$$
$$+ e^{-2nQ\Delta a}\left\{(\Delta a)\left(n^2 Q^3\right) - \frac{5}{2}\, nQ^2\right\} \tag{19a}$$
$$\nu_0 = \left[-3Q - 2\,(1+2an)\,Q^2 + \frac{2K}{n}\right]$$
$$+ e^{-nQ\Delta a}\left\{(\Delta a)^3\left(\frac{7}{6}\, n^3 Q^4\right) + (\Delta a)^2\left(3n^2\, Q^3\right) + (\Delta a)\left(4n\, Q^2\right) + \right.$$
$$\left. + \left[Q^3\left(1 - a^2 n^2\right) + Q^2\left(4an - \frac{1}{4}\right) + \frac{7}{2}\, Q - \frac{2K}{n}\right]\right\} + Q\, e^{-2n\, Q\, \Delta a}$$

$$\tilde{\nu}_3 = \left(\frac{4}{3}\, n^3 Q^4\right) + \left(-\frac{1}{6}\, n^3 Q^4\right) e^{-nQ\,\Delta a}$$
$$\tilde{\nu}_2 = \left(-2n^2 Q^4\right) + \left\{-\frac{1}{2}\, n^2 Q^3\left[nQ\,(a + 7\Delta a) - 4Q + 7\right]\right\} e^{-nQ\,\Delta a}$$
$$\tilde{\nu}_1 = Q\left[\frac{11}{2}\, nQ + 12an^2\, Q^2 + 4nQ^2 - 4K\right] +$$
$$+ e^{-nQ\Delta a}\left\{Q\left[-nQ^2\,(4 + 12an + 5n(\Delta a)) - 5nQ + 4K\right]\right\}$$
$$+ e^{-2nQ\Delta a}\left\{nQ^2\left[Q\,(1 + n(\Delta a - a)) - \frac{7}{2}\right]\right\} \tag{19b}$$
$$\tilde{\nu}_0 = \left(-6anQ^2 - 5Q + \frac{2K}{n}\right) + e^{-nQ\Delta a}\left[6Q^2\,(1 + n\Delta a) - \frac{2K}{n}\right] + Qe^{-2nQ\Delta a}$$

$$\tilde{\tilde{\nu}}_3 = \frac{1}{3}\, n^2 Q^2\,(C_2 + A_2)$$
$$\tilde{\tilde{\nu}}_2 = n^2 Q^2\left(-C_1 - \frac{1}{n}\, C_2 + A_1\right)$$
$$\tilde{\tilde{\nu}}_1 = 2n^2 Q^2\left[\frac{1}{n}\, C_1 + a\left(\frac{a}{2} + \frac{1}{nQ}\right) C_2 - a\, A_1 - \right.$$
$$\left. \frac{1}{2}\left(a^2 + \frac{1}{n^2 Q^2}\right) A_2 - \frac{1}{nQ}\, A_3 + \frac{1}{nQ}\, A_4 + \frac{1}{n^2 Q}\, A_5\right]$$
$$\tilde{\tilde{\nu}}_0 = c_1\left\{2n^2 Q^2\left[-(\Delta a)^2 + a\Delta a - \frac{1}{2}\, a^2\right] + 2nQ\,(\Delta a - a)\right\}$$
$$+ c_2\left\{2n^2 Q^2\left[-\frac{2}{3}(\Delta a)^3 + \frac{1}{2}\, a(\Delta a)^2 - \frac{1}{6}\, a^3\right] + nQ\left((\Delta a)^2 - a^2\right)\right\}$$
$$+ \overline{B}_1\left[2nQ\,(a - 2\Delta a)\right] + \overline{B}_2\left\{2n^2 Q^2\left[-\frac{2}{3}\left((a - \Delta a)^3 - (\Delta a)^3\right) + \right.\right.$$
$$\left.\left. + \frac{1}{2}\, a^2\,(a - 2\Delta a)\right] + nQa\,(a - 2\Delta a)\right\}$$
$$+ A_1\left\{2n^2 Q^2\left[(\Delta a)^2 - a(\Delta a) + \frac{1}{2}\, a^2\right] + 2nQ(\Delta a) - 1\right\}$$

$$+ A_2\left\{2n^2Q^2\left[\frac{2}{3}(a-\Delta a)^3+\frac{1}{2}a\left(2a\,\Delta a-(\Delta a)^2\right)-\frac{1}{3}a^3\right]+\right.$$
$$\left.+ nQ\left(2a\,\Delta a-(\Delta a)^2\right)\right\}$$
$$+ A_3\left\{4e^{-nQ\,\Delta a}\left[nQ(a-\Delta a)-1\right]-2Q\,e^{-nQ\Delta a}+4\right\}$$
$$+ A_4\left\{-4e^{nQ\Delta a}\left[nQ(a-\Delta a)+1\right]+2Qe^{nQ\Delta a}\right\}$$
$$+ A_5\left\{-\frac{4}{nQ}\,e^{nQ\Delta a}\left[n^2Q^2(a-\Delta a)^2+2nQ(a-\Delta a)+2\right]+\right.$$
$$\left.+\frac{2}{nQ}+\frac{2}{n}\,e^{nQ\Delta a}\left[nQ(a-\Delta a)+1\right]\right\}. \tag{19c}$$

The $B_j$ $(j=0,1,2,3)$ are expressible as follows:

$$B_3=\frac{N_3}{D_3},\quad \text{where} \tag{20}$$

$$N_3 = I_0\left\{Q^4\left[a^2n^3\left(-7+\frac{58}{3}\,an-\frac{56}{3}\,a^2n^2+\frac{26}{3}\,a^3n^3-2a^4n^4\right)\right]+\right.$$
$$\left.+Q^2\left[-2an^2+12a^2n^3-\frac{20}{3}\,a^3n^4+2a^4n^5+n\right]+Q\left[4n\right]\right\}$$
$$+I_1\left\{Q^5\left[-4a^2n^4\left(1-\frac{2}{3}\,an\right)\right]+\right.$$
$$+Q^3\left[-4n^2\left(-4an+1+6a^2n^2-\frac{10}{3}\,a^3n^3+a^4n^4\right)\right]+$$
$$\left.+Q^2\left[-4n^2\right]\right\}$$
$$+I_2\left\{Q^5\left[4a^2n^5\right]+Q^4\left[6an^4\right]+Q^3\left[2n^3\right]\right\}$$
$$+I_3\left\{Q^5\left[-\frac{8}{3}\,an^5\right]+Q^4\left[-\frac{4}{3}\,n^4\right]\right\}$$
$$-2nC_1\left\{Q^3\left[a^2n\left(\frac{1}{2}-\frac{2}{3}\,an\right)\right]+\right.$$
$$+Q^2\left[-a^2n\left(3-6an+\frac{10}{3}\,a^2n^2-a^3n^3\right)\right]+$$
$$\left.+Q\left(\frac{3}{2}\,a\right)-\frac{1}{2n}\right\} \tag{21}$$

where the $I_j$ $(j=0,1,2,3)$ are given in the Appendix.
and

$$D_3=\frac{Q^4}{4n^3}\left\{\left[-\frac{28}{15}\,a^7n^7+\frac{58}{15}\,a^6n^6-2a^5n^5\right]+\right.$$
$$+Q^{-2}\left[-14a^8n^8+\frac{218}{3}\,a^7n^7-\frac{548}{3}\,a^6n^6+\frac{1208}{5}\,a^5n^5-154a^4n^4+38a^3n^3\right]$$
$$+Q^{-3}\left[-14a^4n^4+26a^3n^3-12a^2n^2\right]+$$
$$+Q^{-4}\left[12a^6n^6-46a^5n^5+92a^4n^4-68a^3n^3+6a^2n^2+6an\right]+$$

$$+Q^{-5}\left[12a^2n^2-6an\right]+Q^{-6}\left[-6an+3\right]\Big\}. \tag{22}$$

Continuing,

$$B_1=\frac{N_1}{D_1}\quad,\quad\text{with}\quad D_1=D_3\quad,\text{ and} \tag{23}$$

$$N_1=\left(F+G\underline{k}-4an^2Q^2\underline{m}\right)R_1-\left(H+J\underline{k}+L\underline{m}\right)R_2 \tag{24}$$

in which

$$\begin{aligned}
F&=a^2\left(3-a^2n^2Q^2\right)\\
G&=2a\left(3-2a^2n^2Q^2\right)\\
H&=a^3\left[\frac{1}{2}\,a\left(K-3an^2Q^2-nQ\right)-\frac{2}{5}\,a^2n^2Q^2+1\right]\\
J&=a^2\left[2a\left(K-3an^2Q^2-nQ\right)-\frac{3}{2}\,a^2n^2Q^2+3\right]\\
L&=\left[2a\left(K-3an^2Q^2-nQ\right)-a^2n^2Q^2+1\right]\\
\underline{k}&=-\frac{1}{2n}\\
\underline{m}&=-\frac{1}{2n^2Q^2}\left[\frac{3}{2n}\left(1-a^2n^2Q^2\right)-a\left(3-a^2n^2Q^2\right)\right].
\end{aligned} \tag{25}$$

with:

$$\begin{aligned}
R_1=I_0&\left\{Q^4\left[a^2n^3\left(\frac{22}{3}\,an-7\right)\right]+Q^2\left[-10an^2+n\right]+Q\left[2anK-4n\right]+2K\right\}\\
&+I_1\left\{16an^3Q^3+4n^2Q^2-4nQK\right\}+I_2\left\{-2n^3Q^4\right\}+I_3\left\{\frac{4}{3}\,n^4Q^4\right\}
\end{aligned} \tag{26}$$

and

$$R_2=I_0\left\{Q^3\left(-4an^3\right)-4n^2Q^2\right\}+I_1\left\{Q^3\left(8n^3\right)\right\}-2nQ\,C_1. \tag{27}$$

Also,

$$B_2=+3\underline{k}B_3 \tag{28}$$

and

$$B_0=\underline{\ell}+\underline{k}B_1+\underline{m}B_3 \tag{29}$$

where:

$$\underline{\ell}\ =-nQ^2I_0+2n^2Q^2I_1. \tag{30}$$

## 3. Behavior of the Solution

Graphs of $\beta\,K_T^{-1}$ vs. $\ell$ have been plotted (Figs. 1 and 2) for a SW fluid using the simplified form:

$$\begin{aligned}
\beta\,K_T^{-1}=&\frac{Q^2}{\ell}-\frac{2e_1}{\ell}\left(1-e^{-\frac{Q}{\ell}\,\Delta\,a}\right)+\\
&+\frac{2}{\ell^2}\,e_1\left\{\sum_{j=0}^{3}\frac{B_j}{(j+1)}\,a^{j+1}+\sum_{j=0}^{3}\frac{\tilde{\nu}_j}{(j+1)}\,(\Delta\,a)^{j+1}+\right.
\end{aligned}$$

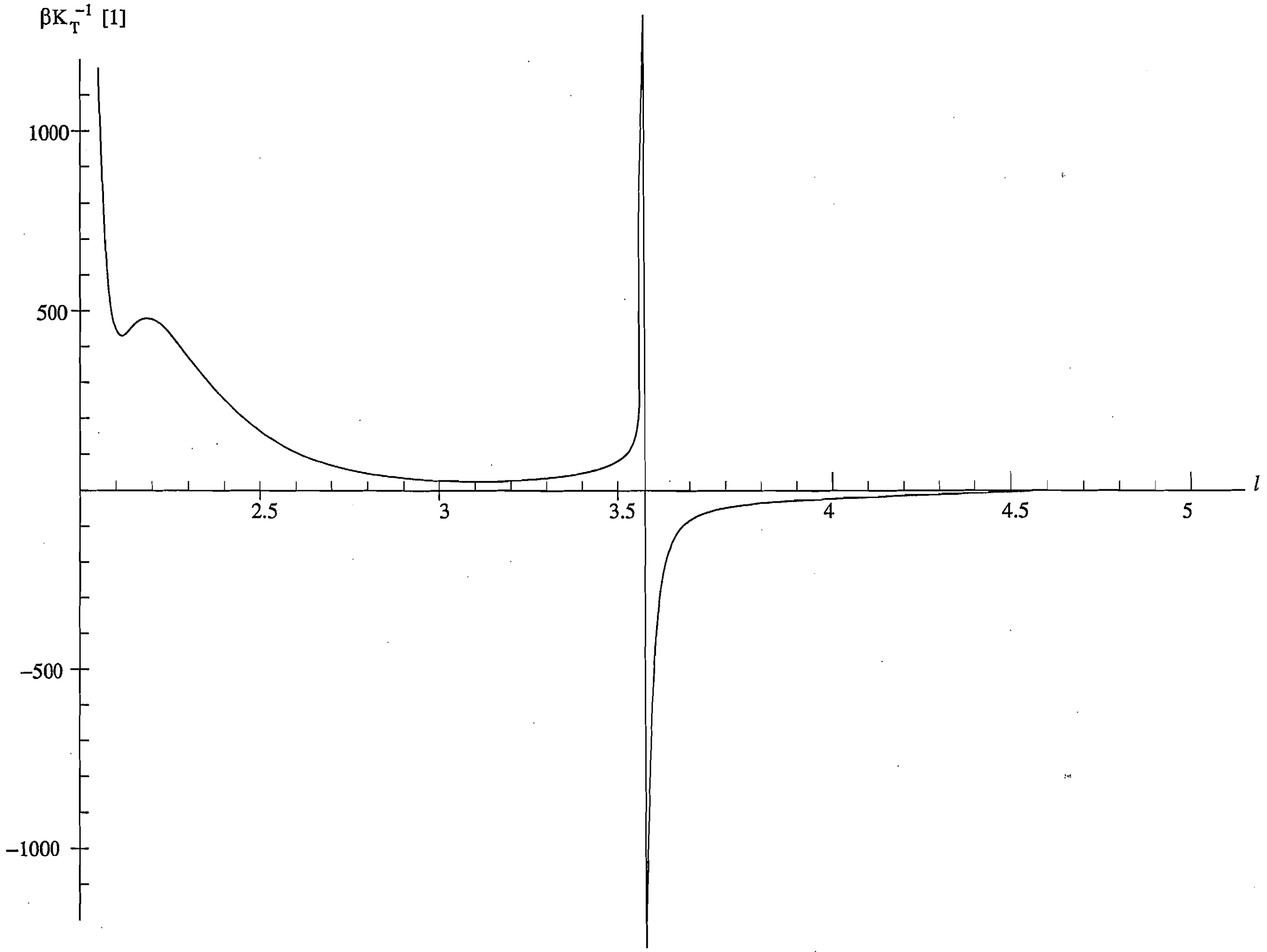

**Figure 1.** Perturbative Plot for $e^{-\beta V_1} = 10$

$$+\sum_{j=0}^{3} \frac{\nu_j}{(j+1)} \left[(a - \Delta a)^{j+1} - (\Delta a)^{j+1}\right]$$
$$+\sum_{j=0}^{3} \frac{\overset{\approx}{\nu}_j}{(j+1)} \left[a^{j+1} - (a - \Delta a)^{j+1}\right]$$
$$\left. +\frac{\ell}{2}\, e^{-\frac{Q}{\ell}\Delta a} + \left[\frac{\ell}{4} - \frac{1}{2}(\Delta a)\, Q\right] e^{-2\frac{Q}{\ell}\Delta a} - \frac{3}{4}\ell \right\} \qquad (31)$$

where: $Q = \frac{\ell}{\ell - a}$, and $e_1 = \left(e^{-\beta V_1} - 1\right)$.

The parameters for the SW potential are the core diameter $a$, the well width $\Delta a$, and the well depth $V_1$. Clearly, $1 < \exp(-\beta V_1) < \infty$. A shallower well ($\exp(-\beta V_1) = 10$), (see Fig. 1) and a deeper well ($\exp(-\beta V_1) = 100$) (see Fig. 2) were graphed for $a = 2.00$ and $\Delta a = 0.500$. When $\Delta a$ or $V_1$ is varied, a singularity again occurs at the same location, $\ell_0 = 3.5465611\ldots$ $\Delta a$ and $V_1$ have no effect on the location of the singularity at $\ell_0$, but will affect the thermodynamics. From equation (22) for $D_3[\ell]$, the value of $D_3[\ell]$ becomes zero between $\ell = 3.5465611$ and $3.5465612$, causing

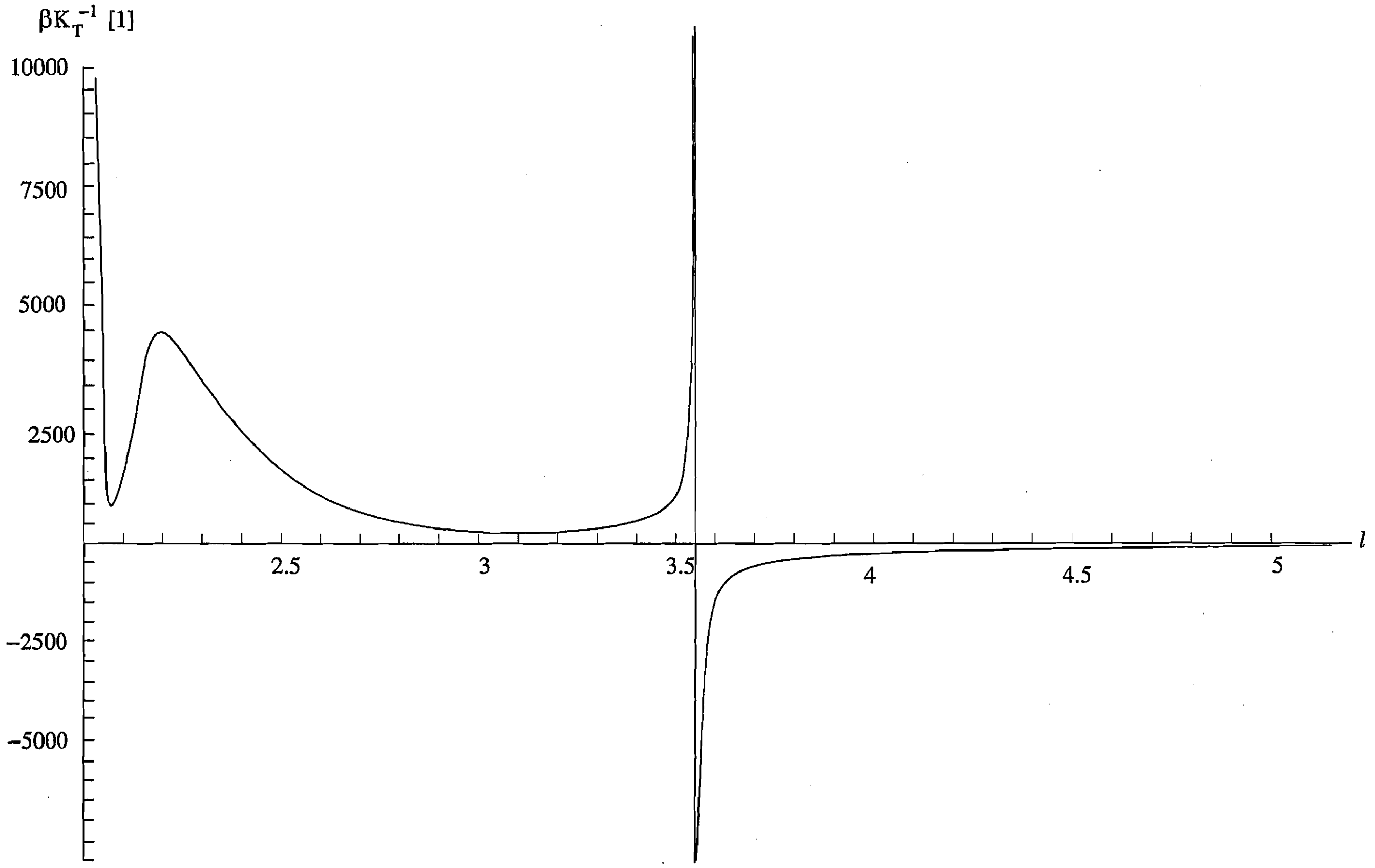

**Figure 2.** Perturbative Plot for $e^{-\beta V_1} = 100$

$B_0, B_1, B_2$ and $B_3$ to be singular at $\ell_0$. (See Fig. 3). The ratio $a/\ell_0$ for which $D_3$ vanishes is always the same.

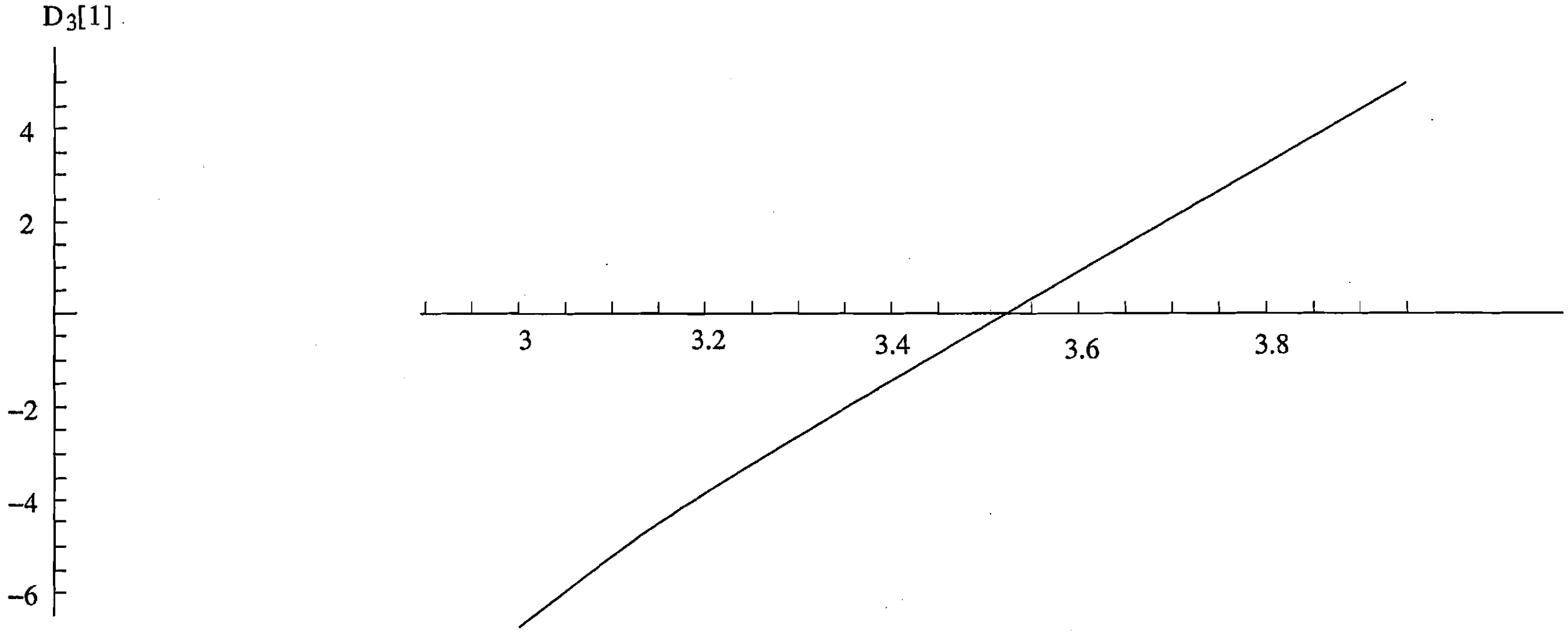

**Figure 3.** Plot of $D_3$ v. 1

## 4. Analysis of the Singularity

It is difficult to obtain physical insight into the singularity. However, for a first-order perturbation, terms linear in $(\Delta\, a)$ must be responsible. Terms of zeroth order in $(\Delta a)$ have no effect and will disappear, i.e., the coefficient of $(\Delta a)^0$ in $N_3$ is zero, as may readily be shown. Terms of higher order in $(\Delta a)$ are not considered.

For a first-order singular perturbation, the coefficient of $(\Delta a)$ in $N_3$ will not be zero. The calculation is straightforward. Referring to equation (21),

coefficient of $(\Delta a)$ in $N_3$ due to $I_0$

$$= \left(2an + 5a^2 n^2\right) Q^3 - 2aKQ \tag{32a}$$

coefficient of $(\Delta a)$ in $N_3$ due to $I_1$

$$= \left(-a^2 KQ + a^2 nQ^3 + \frac{8}{3}\, a^3 n^2 Q^3\right) \tag{32b}$$

coefficient of $(\Delta a)$ in $N_3$ due to $I_2$

$$= \left(-\frac{2}{3}\, a^3 KQ + \frac{2}{3}\, a^3 nQ^3 + \frac{11}{6}\, a^4 n^2 Q^3\right) \tag{32c}$$

coefficient of $(\Delta a)$ in $N_3$ due to $I_3$

$$= \left[\left(\frac{1}{2}\, a^4 n + \frac{7}{5}\, a^5 n^2\right) Q^3 - \frac{1}{2}\, a^4 KQ + \frac{12}{n^3 Q^2}\right] \tag{32d}$$

coefficient of $(\Delta a)$ in $N_3$ due to $-2n\, C_1$

$$= -2n \left(6an^2 Q^3 - 2KQ\right) \tag{32e}$$

The desired coefficient of $(\Delta a)$ in $N_3$ can be assembled using (32*a*) through (32*e*). Since only $I_0, I_1, I_2, I_3$, and $-2n\, C_1$ contain $(\Delta a)$ (refer to eqn. 21), there follows

$$\begin{aligned}
N_3/(\Delta a) = {} & Q^8 \left(a^4 n^5\right) \left(-4 - \frac{20}{3}\, an + \frac{482}{45}\, a^2 n^2\right) \\
& + Q^7 \left(a^3 n^4\right) \left(-14 + 7an + \frac{1027}{15}\, a^2 n^2 - 76a^3 n^3 + \frac{118}{3}\, a^4 n^4 - 10a^5 n^5\right) \\
& + Q^6 \left(a^2 n^3\right) \left(-4 + \frac{2}{3}\, an + \frac{91}{3}\, a^2 n^2 - \frac{152}{3}\, a^3 n^3 + \frac{284}{9}\, a^4 n^4 - \right. \\
& \qquad \left. - \frac{32}{3}\, a^5 n^5 + 4a^2 nK \left(1 - an\right)\right) \\
& + Q^5 \left\{\left(2an^2\right) \left[1 - 2an + 12a^2 n^2 - \frac{20}{3}\, a^3 n^3 + 2a^4 n^4\right] + \right. \\
& \qquad \left. + 2a^3 n^3 K \left[7 - \frac{58}{3}\, an + \frac{56}{3}\, a^2 n^2 - \frac{26}{3}\, a^3 n^3 + 2a^4 n^4\right]\right\} \\
& + Q^4 \left\{2an^2 \left(4 + an\right) + 2a^2 n^2 K \left[3 - 10an + 12a^2 n^2 - \frac{20}{3}\, a^3 n^3 + 2a^4 n^4\right]\right\} \\
& + Q^3 \left\{-26an^2 - 2an \left(1 + 2an\right) K\right\} \\
& + Q^2 \left\{-16n - 2an\, K\right\} + Q \left\{-2K\right\}.
\end{aligned} \tag{33}$$

The quantity $N_3/(\Delta a)$ represents the coefficient of $(\Delta a)$ in $N_3$.

## 5. Conclusions

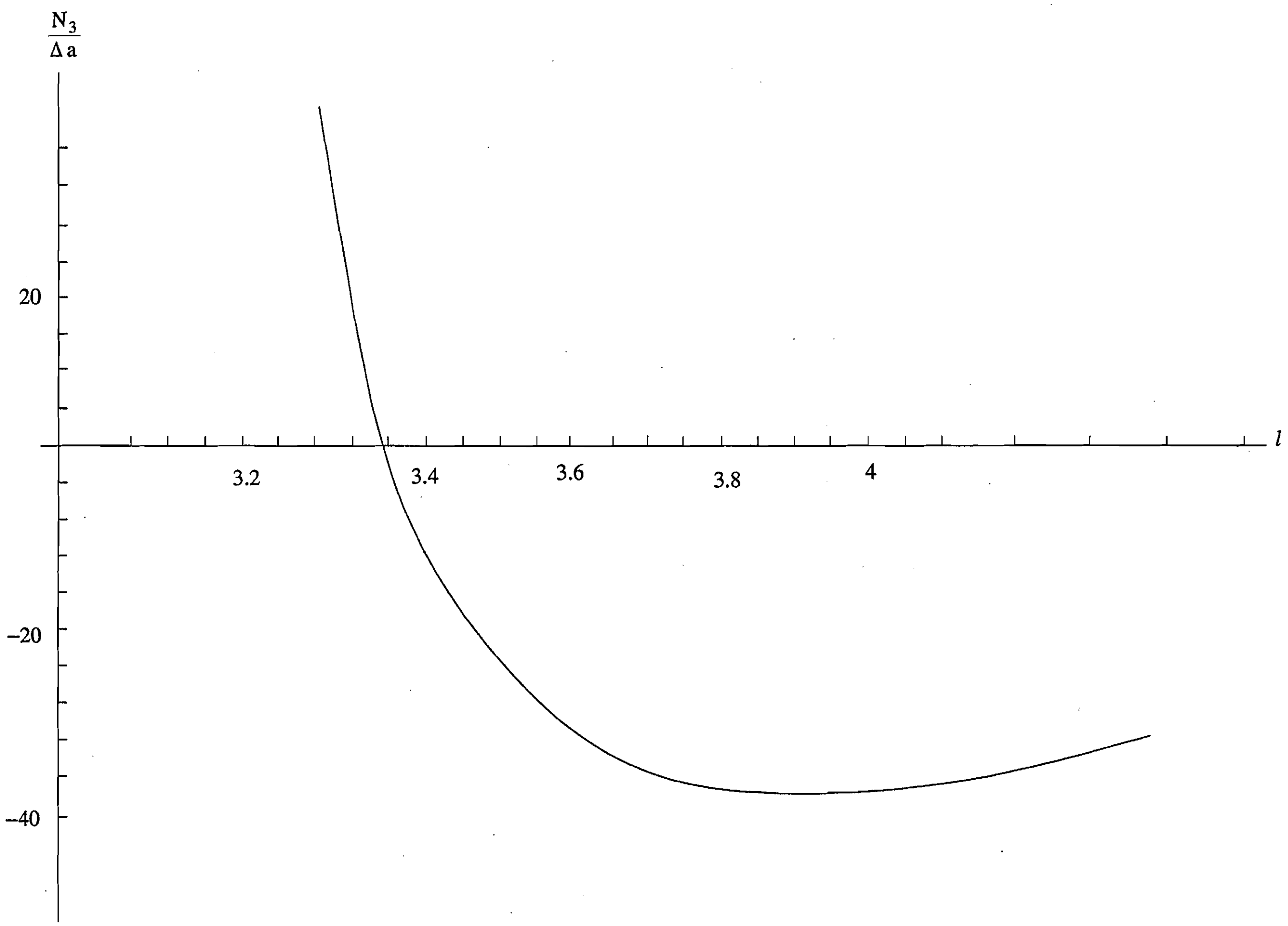

**Figure 4.** Detail of Plot $\frac{N_3}{\Delta a}$ vs. 1

The graph of equation (33) for $3 < \ell < 4$ (see Fig. 4) crosses the axis at $\ell = 3.34012\ldots$, not equal to $\ell_0$, where the singularity occurs. Thus, there is confirmation for a first-order singular perturbation for a SW fluid.

Whether a singularity in $\beta\, K_T^{-1}$ occurs at a characteristic density in other first-order equations for the pair correlation function has not been investigated.

## Acknowledgments

The author is greatly indebted to Professor J. K. Percus for significant scholarly assistance and for encouragement which allowed the work to be completed. He is grateful to Professor Rodney Varley for helpful advice and discussions, and to Charlotte Sanders for invaluable support with the coding and programming.

## Appendix

Where the $I_j$ $(j = 0, 1, 2, 3)$ are given in the Appendix:

$$\begin{aligned} I_0 = \Big[ & c_1\,(\Delta a) + \frac{1}{2}\,c_2\,(\Delta a)^2 + \overline{B}_1\,(a - 2\Delta a) + \frac{1}{2}\,\overline{B}_2\left(a^2 - 2a\,\Delta a\right) + \\ & + A_1\,(\Delta a) - \frac{1}{2}\,A_2\left(-2a\,\Delta a + (\Delta a)^2\right)\Big] + \\ & + \frac{A_3}{nQ}\left(1 - e^{-nQ\,\Delta a}\right) - \frac{A_4}{nQ}\left(1 - e^{nQ\,\Delta a}\right) + \\ & + \frac{A_5}{n^2\,Q^2}\left[-(nQa + 1) + e^{nQ\,\Delta a}\,(nQ(a - \Delta a) + 1)\right]. \end{aligned} \tag{A.1}$$

$$\begin{aligned} I_1 = \Big\{ & \frac{1}{2}\,c_1(\Delta a)^2 + \frac{1}{3}\,c_2(\Delta a)^3 + \frac{1}{2}\,\overline{B}_1\left[(a - \Delta a)^2 - (\Delta a)^2\right] + \\ & + \frac{1}{3}\,\overline{B}_2\left[(a - \Delta a)^3 - (\Delta a)^3\right] + \frac{1}{2}\,A_1\left[a^2 - (a - \Delta a)^2\right] + \\ & + \frac{1}{3}\,A_2\left[a^3 - (a - \Delta a)^3\right]\Big\} + \\ & + \frac{A_3}{n^2\,Q^2}\left[(nQa - 1) - e^{-nQ\,\Delta a}\,(nQ(a - \Delta a) - 1)\right] + \\ & + \frac{A_4}{n^2\,Q^2}\left[-(nQa + 1) + e^{nQ\,\Delta a}\,(nQ(a - \Delta a) + 1)\right] - \\ & - \frac{A_5}{n^3\,Q^3}\Big[\left(n^2\,Q^2\,a^2 + 2n\,Qa + 2\right) - \\ & - e^{nQ\,\Delta a}\left(n^2\,Q^2(a - \Delta a)^2 + 2nQ(a - \Delta a) + 2\right)\Big] \end{aligned} \tag{A.2}$$

$$\begin{aligned} I_2 = \Big\{ & \frac{1}{3}\,c_1\,(\Delta a)^3 + \frac{1}{4}\,c_2\,(\Delta a)^4 + \frac{1}{3}\,\overline{B}_1\left[(a - \Delta a)^3 - (\Delta a)^3\right] + \\ & + \frac{1}{4}\,\overline{B}_2\left[(a - \Delta a)^4 - (\Delta a)^4\right] + \\ & + \frac{1}{3}\,A_1\left[a^3 - (a - \Delta a)^3\right] + \frac{1}{4}\,A_2\left[a^4 - (a - \Delta a)^4\right]\Big\} \\ & + \frac{A_3}{n^3\,Q^3}\Big[\left(n^2\,Q^2\,a^2 - 2nQa + 2\right) - \\ & - e^{-nQ\,\Delta a}\left(n^2\,Q^2(a - \Delta a)^2 - 2nQ\,(a - \Delta a) + 2\right)\Big] \\ & - \frac{A_4}{n^3\,Q^3}\Big[\left(n^2\,Q^2\,a^2 + 2nQa + 2\right) - \\ & - e^{nQ\,\Delta a}\left(n^2\,Q^2\,(a - \Delta a)^2 + 2nQ\,(a - \Delta a) + 2\right)\Big] \\ & - \frac{A_5}{(nQ)^4}\Big\{\left[(nQa)^3 + 3(nQa)^2 + 6(nQa) + 6\right] - \\ & - e^{nQ\,\Delta a}\Big[(nQ(a - \Delta a))^3 + 3\,(nQ\,(a - \Delta a))^2 + \\ & + 6\,(nQ\,(a - \Delta a)) + 6\Big]\Big\}. \end{aligned} \tag{A.3}$$

$$
\begin{aligned}
I_3 = \Big\{ &\frac{1}{4}\, c_1(\Delta a)^4 + \frac{1}{5}\, c_2(\Delta a)^5 + \frac{1}{4}\, \overline{B}_1 \left[(a-\Delta a)^4 - (\Delta a)^4\right] + \\
&+\frac{1}{5}\, \overline{B}_2 \left[(a-\Delta a)^5 - (\Delta a)^5\right] + \frac{1}{4}\, A_1 \left[a^4 - (a-\Delta a)^4\right] + \\
&+\frac{1}{5}\, A_2 \left[a^5 - (a-\Delta a)^5\right]\Big\} \\
&+\frac{A_3}{(nQ)^4} \Big\{\left[(nQa)^3 - 3(nQa)^2 + 6(nQa) - 6\right] - \\
&-e^{-nQ\,\Delta a} \left[(nQ(a-\Delta a))^3 - 3\,(nQ(a-\Delta a))^2 + 6\,(nQ(a-\Delta a)) - 6\right]\Big\} \\
&-\frac{A_4}{(nQ)^4} \Big\{\left[(nQa)^3 + 3\,(nQa)^2 + 6\,(nQa) + 6\right] - \\
&-e^{nQ\,\Delta a} \left[(nQ(a-\Delta a))^3 + 3\,(nQ(a-\Delta a))^2 + 6\,(nQ(a-\Delta a)) + 6\right]\Big\} \\
&-\frac{A_5}{(nQ)^5} \Big\{\left[(nQa)^4 + 4(nQa)^3 + 12(nQa)^2 + 24(nQa) + 24\right] - \\
&-e^{nQ\,\Delta a} \Big[(nQ(a-\Delta a))^4 + 4\,(nQ(a-\Delta a))^3 + \\
&+12\,(nQ(a-\Delta a))^2 + 24\,(nQ(a-\Delta a)) + 24\Big]\Big\}\,. \qquad \text{(A.4)}
\end{aligned}
$$